\documentclass{vgtc}                          

\newcommand{\eat}[1]{}

\usepackage{latexsym}
\usepackage{amsfonts}
\usepackage{amsmath}
\usepackage{amssymb}
\usepackage{xcolor}
\usepackage{colortbl}
\usepackage{epsfig}
\usepackage{xspace}
\usepackage{graphicx}
\usepackage{subfigure}
\usepackage{paralist}
\usepackage{enumerate}
\usepackage[color,matrix,arrow,all]{xy}
\usepackage{comment}
\usepackage{booktabs}
\usepackage{balance}
\usepackage{stmaryrd}
\usepackage{pifont}
\usepackage{hhline}
\usepackage{listings}
\usepackage{array}
\usepackage{float}
\usepackage[flushleft]{threeparttable}

\usepackage{wrapfig}

\usepackage{epsfig}
\usepackage{multirow}
\usepackage{url}

\definecolor{inputcolor}{RGB}{255,139,35}
\definecolor{outputcolor}{RGB}{120,212,252}
\definecolor{embedcolor}{RGB}{254,127,156}
\definecolor{maskcolor}{RGB}{122,128,255}
\definecolor{ecolor}{RGB}{58,149,54}

\newcommand{\encodec}[1]{\textcolor{inputcolor}{\att{#1}}}
\newcommand{\decodec}[1]{\textcolor{outputcolor}{\att{#1}}}

\usepackage{tikz}
\usetikzlibrary{shapes,snakes}
\usetikzlibrary{calc}

\usepackage[ruled,vlined,linesnumbered]{algorithm2e}
\usepackage[export]{adjustbox}

\newcommand{\att}[1]{\textbf{\texttt{#1}}}

\newcommand{\nlp}{\texttt{NL}\xspace}

\newcommand{\sql}{\texttt{SQL}\xspace}
\newcommand{\vis}{\texttt{VIS}\xspace}
\newcommand{\nlsql}{\texttt{NL2SQL}\xspace}
\newcommand{\nlvis}{\texttt{NL2VIS}\xspace}
\newcommand{\sys}{\att{ncNet}\xspace}

\newcommand{\tvbench}{\att{nvBench}\xspace}
\newcommand{\seqseq}{\texttt{seq2seq}\xspace}
\newcommand{\deepeye}{\texttt{DeepEye}\xspace}
\newcommand{\x}{\mathbf{x}}

\newcommand{\bi}{\begin{itemize}}
\newcommand{\ei}{\end{itemize}}

\newcommand{\be}{\begin{enumerate}}
\newcommand{\ee}{\end{enumerate}}
\newcommand{\beqn}{\begin{eqnarray*}}
\newcommand{\eeqn}{\end{eqnarray*}}

\newcommand{\stitle}[1]{\vspace{1ex}\noindent{\bf #1}}

\newcommand{\ie}{{\em i.e.,}\xspace}
\newcommand{\eg}{{\em e.g.,}\xspace}
\newcommand{\wrt}{\emph{w.r.t.}\xspace}

\makeatletter
    \newcommand\figcaption{\def\@captype{figure}\caption}
    \newcommand\tabcaption{\def\@captype{table}\caption}
\makeatother

\tikzstyle{mybox} = [draw=black, fill=black!5, thick,
   rectangle, rounded corners, inner sep=0pt, inner ysep=6pt]
\tikzstyle{fancytitle} =[fill=black, text=white]

\ifpdf
  \pdfoutput=1\relax                   
  \usepackage{graphicx}                
  \DeclareGraphicsExtensions{.pdf,.png,.jpg,.jpeg} 
\else
  \usepackage{graphicx}                
  \DeclareGraphicsExtensions{.eps}     
\fi%

\graphicspath{{figures/}{pictures/}{images/}{./}} 

\usepackage{microtype}                 
\PassOptionsToPackage{warn}{textcomp}  
\usepackage{textcomp}                  
\usepackage{mathptmx}                  
\usepackage{times}                     
\usepackage{cite}                      
\usepackage{tabu}                      
\usepackage{booktabs}                  

\onlineid{0}

\vgtccategory{Research}

\vgtcinsertpkg

\title{nvBench: A Large-Scale Synthesized Dataset for Cross-Domain \\ Natural Language to Visualization Task}

\author{Yuyu Luo, Jiawei Tang, Guoliang Li}
\authorfooter{
	\item
	Yuyu Luo and Guoliang Li are with  the Department of Computer Science, Tsinghua University, China. Email: \{luoyy18@mails., liguoliang@\}tsinghua.edu.cn
	\item
	Jiawei Tang is with American School of Doha, Doha, Qatar. Email: 23jtang@asd.edu.qa
}

\author{Yuyu Luo\\ %
	\scriptsize  luoyy18@mails.tsinghua.edu.cn \\
    \scriptsize Tsinghua University, China %
\and Jiawei Tang\\ %
\scriptsize 23jtang@asd.edu.qa \\
    \scriptsize American School of Doha, Qatar %
\and Guoliang Li\\ %
\scriptsize liguoliang@tsinghua.edu.cn \\
    \scriptsize Tsinghua University, China %
}

\teaser{
	\hspace{-2.5em}
	\includegraphics[width=1.1\textwidth]{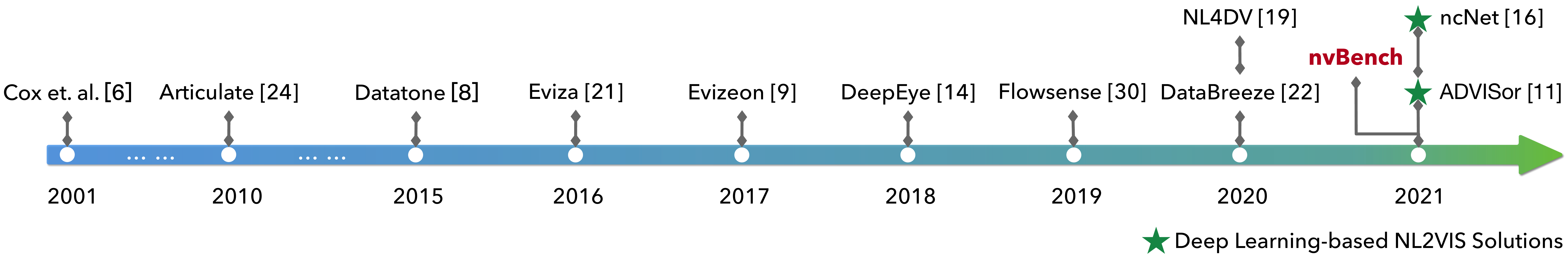}	
	\vspace{-2em}
	\caption{A brief history of natural language query to visualization in academia. After the release of \tvbench in 2021, some deep learning-based models are developed to support translating natural language queries into visualizations.}
	\label{fig:history}
}

\abstract{
  \nlvis~-- which translates natural language (\nlp) queries to corresponding visualizations (\vis) -- has attracted more and more attention both in commercial visualization vendors and academic researchers. 
  In the last few years, the advanced deep learning-based models have achieved human-like abilities in many natural language processing (NLP)  tasks, which clearly tells us that the deep learning-based technique is a good choice to push the field of \nlvis. 
  However, a big balk is the lack of benchmarks with lots of  (\encodec{\nlp}, \decodec{\vis}) pairs. 
  We present \tvbench, the first large-scale \nlvis benchmark, containing 25,750 (\encodec{\nlp}, \decodec{\vis}) pairs from 750 tables over 105 domains, synthesized from (\encodec{\nlp}, \decodec{\sql}) benchmarks to support cross-domain \nlvis task. 
  The quality of \tvbench has been extensively validated  by 23 experts and 300+ crowd workers. 
  Deep learning-based models training using \tvbench demonstrate  that \tvbench can push the field of \nlvis.
  
} 

\begin{document}

\setcounter{figure}{0}
\setcounter{table}{0}
\pagestyle{plain}
\pagenumbering{arabic}

\firstsection{Introduction}
\label{sec:intro}
\maketitle

\begin{figure*}[!htb]
	\centering
	\begin{minipage}{.59\textwidth}
		\includegraphics[width=\columnwidth]{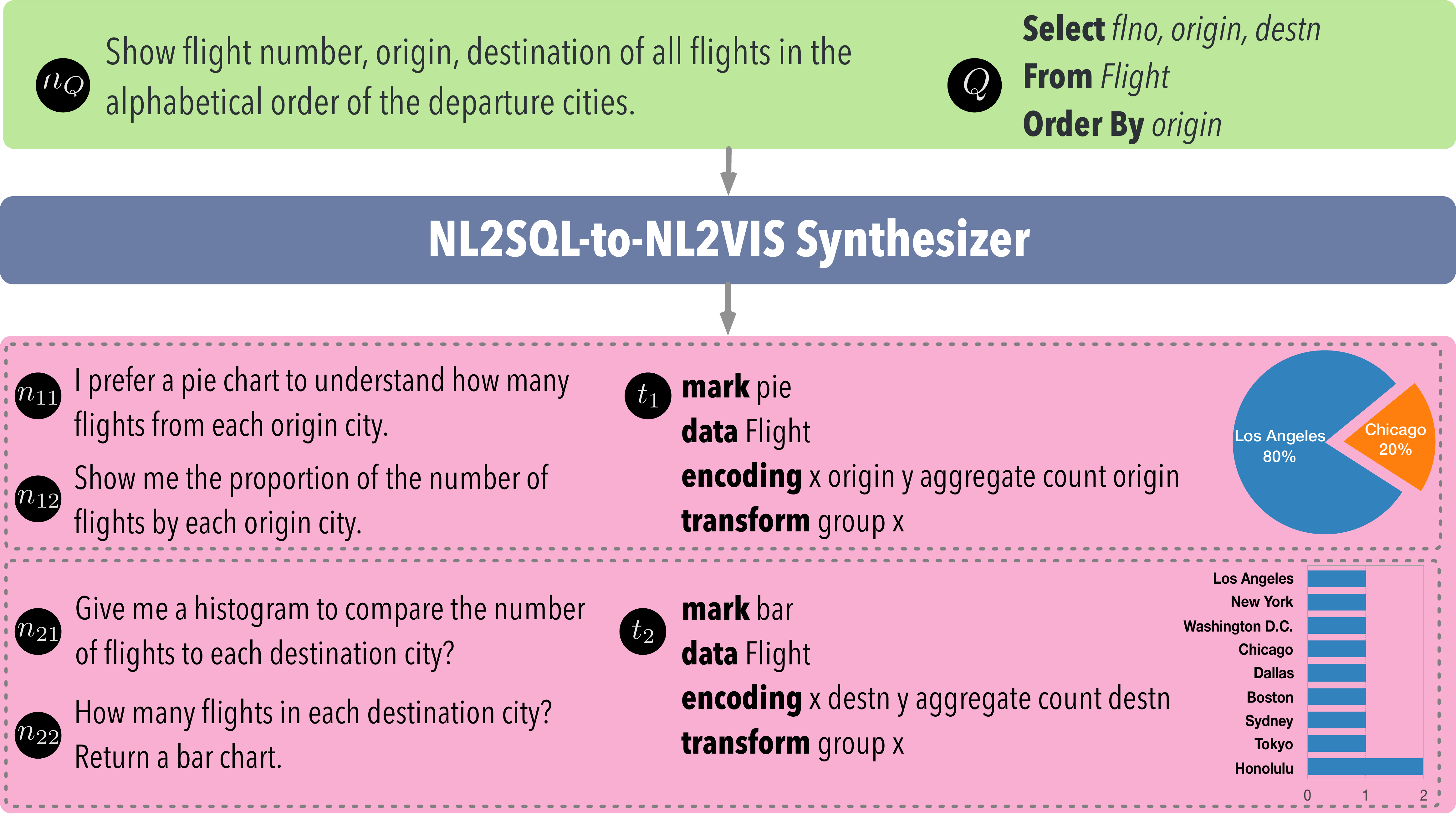}	
		\vspace{-2em}
		\caption{An example of synthesizing multiple (\encodec{\nlp}, \decodec{\vis}) pairs from one (\encodec{\nlp}, \decodec{\sql}) pair}
		\label{fig:synthesizer}
	\end{minipage}%
	\hspace{1em}
	\begin{minipage}{0.275\textwidth}
		\includegraphics[width=\columnwidth]{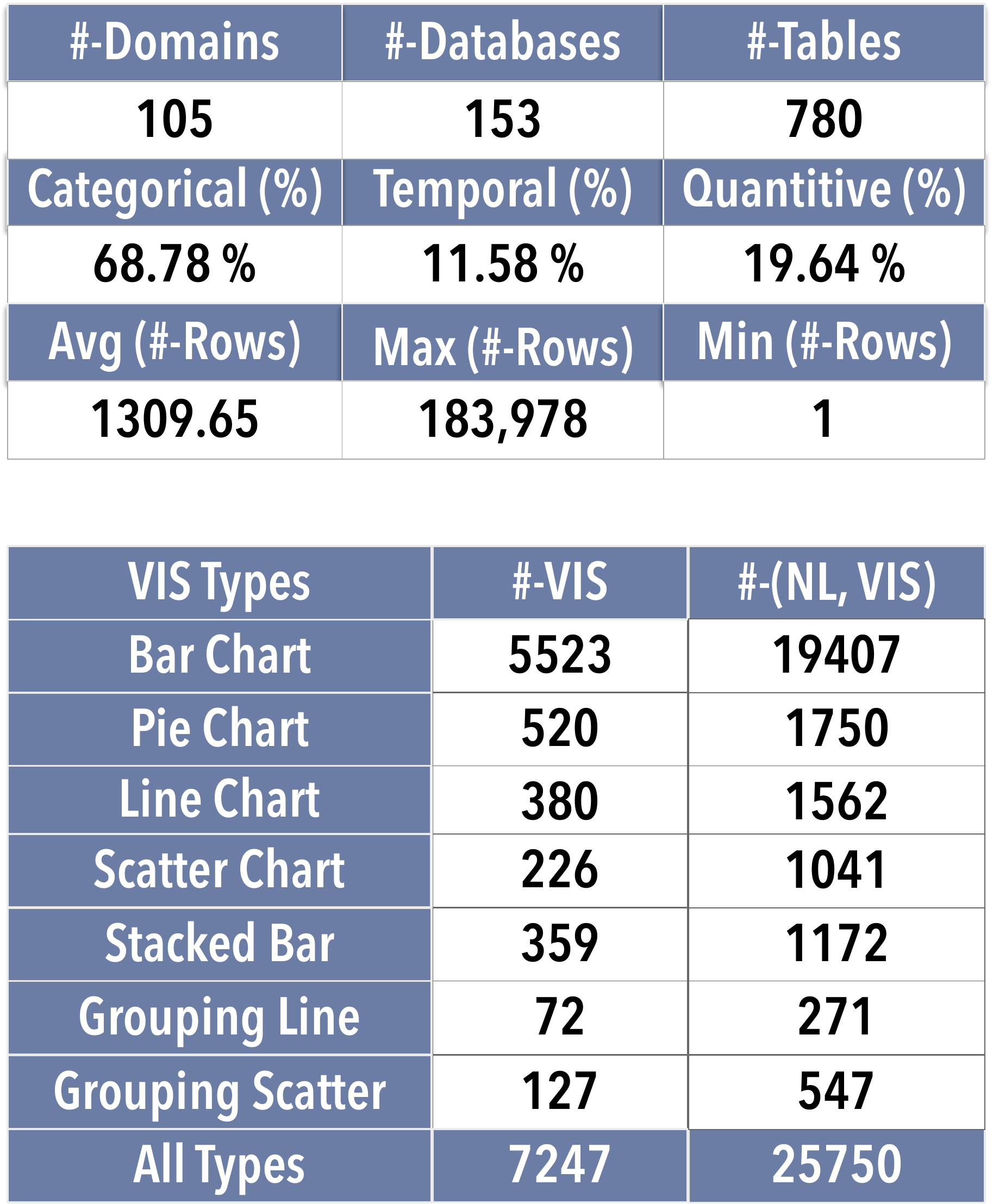}	
		\vspace{-2em}
		\caption{The statistics of \tvbench
		}
		\label{fig:nvbench_stat}
	\end{minipage}
\vspace{-1em}
\end{figure*}

%
%
%
%

Creating meaningful visualizations from data through natural language is a promising interaction paradigm, especially for novices, and is an essential step towards democratizing data visualization~\cite{DBLP:conf/sigmod/TangWL19, DBLP:journals/vldb/QinLTL20, DBLP:conf/icde/LuoCQ0020, DBLP:journals/pvldb/LuoCQ0020, DBLP:journals/pvldb/LuoLZYZ0020, DBLP:journals/debu/Luo00LZY20}. 
Both mainstream commercial vendors (\eg Tableau’s Ask Data~\cite{askdata}, Microsoft Power BI~\cite{msQA}, ThoughtSpot~\cite{spotiq}, and Amazon’s QuickSight~\cite{AMqs}) and academic researchers~\cite{nl4dv,articulate, databreeze, datatone, eviza, evizeon, deepeye_edbt, deepeye_sigmod, deepeye_tkde, deepeye_icde, flowsense, ADVISor, ncnet, cox2001multi} have started to explore the techniques to support \nlvis for decades.

Despite the importance of \nlvis, the study of \nlvis is still in its infancy~\cite{spotiq}. 
Currently, most of the \nlvis systems are developed mainly based on  NLP statistical parsers and only support simple or constrained \nlp queries.
Although cutting-edge deep learning-based models have human-like abilities in many NLP tasks (e.g., text classification, language translation),  such techniques are not equipped to the field of  \nlvis. The main obstacle is that it lacks large-scale and high-quality benchmarks for supporting the \nlvis task, and our goal is to fill this gap.

Given a table (or database), \nlvis can be viewed as a \textit{machine translation} task that translates natural language queries (\eg show me the trend of COVID-19 total confirmed cases in China) to visualization queries (\eg \underline{mark} line \underline{data} COVID-19 \underline{encoding} x date y aggregate sum confirmed \underline{transform} filter country = `China' group~x) so as to be rendered as a visualization specification (\eg Vega-Lite). 
The key factor to making this a success is to acquire enough high-quality (\encodec{\nlp},~\decodec{\vis}) pairs because deep learning models require the availability of large-scale and high-quality training data.

In this paper, we present such a benchmark, namely \tvbench~\cite{nvbench}, that contains 25,750  (\encodec{\nlp}, \decodec{\vis}) pairs over 750 tables from 105 domains to support the cross-domain \nlvis task.  
Different from the common practice that builds such a benchmark by manually designing and collecting enough data and queries,  we synthesize \tvbench by piggybacking \nlsql benchmarks. The intuition is based on the semantic connection between \sql queries and \vis queries: \sql queries specify \textit{what} data is needed and \sql queries additionally need to specify \textit{how} to visualize the data. 
The quality of \tvbench has been validated by experts and crowd-workers, and  a deep learning-based model, namely \sys~\cite{ncnet}, training using \tvbench, also verifies the power of \tvbench.

\section{Related Work}

%
As shown in Figure~\ref{fig:history},  
there has been a surge of works on developing techniques to support translating natural languages to visualizations~\cite{articulate, datatone, eviza, evizeon, deepeye_sigmod, flowsense, nl4dv, databreeze, ncnet, ADVISor, cox2001multi}.

%
\stitle{Rule-based NL2VIS Approaches.}
The idea of creating visualizations using \nlp queries was explored around two decades ago~\cite{cox2001multi}.
Afterward, semantic parser techniques (\eg Stanford Core NLP Parser~\cite{corenlp}) are becoming increasingly popular in the research of \nlvis because these techniques can extract useful semantic information from the \nlp query. 
The Articulate~\cite{articulate} is an \nlvis system that translates the user-provided \nlp query into a representative visualization with two steps. First,  it maps the \nlp query into a set of explicit commands using Stanford Parser and classifies the \nlp query to a set of user tasks using a supervised learning method. Second, it deploys a heuristic algorithm to generate a suitable visualization based on the commands  and data properties automatically.
DataTone~\cite{datatone} mainly utilizes Stanford Core NLP Parser~\cite{corenlp} and a set of rules to mapping an \nlp query into a visualization. It also develops a mixed-initiative method to handle ambiguity in the process of \nlvis.  The user can interact with the ambiguity widgets in the interface to handle the ambiguities.
Eviza~\cite{eviza} is an \nlvis system that  allows users to have a conversation on a given visualization. Eviza develops a probabilistic grammar-based approach and a finite state machine to manage the interaction processing of the \nlvis task. Eviza also manages syntactic and semantic ambiguity through simple GUI widgets in the interface, similar to DataTone~\cite{datatone}.
Evizeon~\cite{evizeon}, extending Eviza's features and introduces additional  pragmatics concepts, enables users to issue standalone and follow-up \nlp queries to specify a new visualization or interact with an existing visualization.
Note that, Ask Data~\cite{askdata} in Tableau is partially based on their previous studies – Eviza~\cite{eviza} and Evizeon~\cite{evizeon}.
DeepEye~\cite{deepeye_sigmod} demonstrates a simple rule-based method for generating \vis charts from (constrained) keyword queries.
Flowsense~\cite{flowsense} uses state-of-the-art semantic parser techniques to support \nlp queries in a dataflow system, which allows users  to use \nlp query for the majority of dataflow diagram editing operations.
NL4DV~\cite{nl4dv} is a Python toolkit that supports to generate data visualization using \nlp queries, mainly based on the NLP parser tree techniques, similar to the previous works (\eg DataTone~\cite{datatone} and Flowsense~\cite{flowsense})

\stitle{NL2VIS Benchmarks.}
A recent work~\cite{nl4vis} collected 893 \nlp queries over 3 datasets by 
conducting an online study with 102 participants. 
This work characterizes the \nlp queries based on the phrasing (\eg what types of keywords are used by real users) and the information contained (\eg aggregation). 
Therefore, the 893 \nlp queries can be used to evaluate the performance of existing \nlvis systems, or used by the developers to design their \nlvis techniques, especially for rule-based techniques. 
However, this dataset has two limitations. 
First, the size of the dataset is too small for training the data-intensive deep learning model. 
Second, since this dataset is curated from 3 tables, it is hard to generalize to real-world scenarios.

Therefore, it needs a large-scale, high-quality, and realistic \nlvis dataset for the cross-domain \nlvis task.


\stitle{Deep learning-based NL2VIS Approaches.}
The aforementioned studies are mainly developed based on rule-based NLP methods, which do not support well in free-form \nlp input. 
Some researchers try to support \nlvis  by applying deep learning-based NLP techniques such as language representation.

ADVISor~\cite{ADVISor}, a deep learning-based pipeline, aims to create visualization relevant to the user-provided \nlp query. 
Roughly speaking, The whole pipeline of ADVISor can be divided into two steps: (1) \nlsql step, and (2) rule-based visualization generation step.
For the first step -- \nlsql , ADVISor uses WikiSQL~\cite{DBLP:journals/corr/abs-1709-00103}, a large crowd-sourced dataset for \nlsql task, as the training dataset.
In this step,
ADVISor firstly takes as input the \nlp query and table headers to a BERT model~\cite{bert}. Next, two neural networks are trained to classify {\em aggregation types}, and {\em relevant attributes and filter conditions}. 
In the second step, ADVISor designs a rule-based method to automatically create a visualization based on the selected \textit{attribute}, \textit{filter conditions}, and \textit{aggregation type}. 
Hence, the neural network components of ADVISor are trained to produce fragments of SQL queries from the given \nlp query.  
It means that the deep learning models of ADVISor do not directly generate the visualization results from the given \nlp query.

Thanks to the large number of (\encodec{\nlp}, \decodec{\vis}) pairs in \tvbench, developers can use these pairs to train an end-to-end neural network for the \nlvis task.
\sys~\cite{ncnet} is a Transformer-based model for translating \nlp queries into visualizations.  It takes  \tvbench as the training corpus  to solve the \nlvis task in an end-to-end way.


\section{Synthesizing \tvbench from NL2SQL Benchmarks}
\label{label: nvbench}

The widely used practice of producing benchmarks is through time-consuming manual labeling, \eg providing visualizations and ask experts to write corresponding \nlp queries.

The main issue of the above approach is that the required experts are simply not enough.
Alternatively, we propose to synthesize \nlvis benchmarks~\cite{nvbench} from a plethora of \nlsql benchmarks.
Because it is known that verifying results (\ie whether an \nlp query is suitable for a given visualization) is much easier than writing the \nlp query manually, both experts and crowd-workers can help.

The rationality that \nlvis benchmarks can be synthesized from \nlsql benchmarks is because of the semantic connection between \vis queries and \sql queries: 
\sql queries specify {\em what} data is needed (\eg columns, filtering, aggregation, sorting);  
and \vis queries specify both {\em what} data is needed and {\em how} to visualize (\eg bar or line charts) -- the {\em what data} parts highly overlap. Intuitively, we can piggyback \nlsql benchmarks on the {\em what data} part and focus on synthesizing {\em how to visualize} for \nlvis.

Briefly speaking, given a (\encodec{\nlp}, \decodec{\sql}) pair, our method will synthesize a set of (\encodec{\nlp}, \decodec{\vis}) pairs.
Consider Figure~\ref{fig:synthesizer}, the input is a pair $(n_Q, Q)$. It outputs four pairs $(v_1, n_{11})$, $(v_1, n_{12})$, $(v_2, n_{21})$, and $(v_2, n_{22})$, where $v_1$ (resp. $v_2$) is a pie (resp. bar) chart, and $n_{11}$ and $n_{12}$ (resp. $n_{21}$ and $n_{22}$) are variants of \nlp queries for $v_1$ (resp. $v_2$).

The synthesis steps from one (\encodec{\nlp}, \decodec{\sql}) pair to multiple (\encodec{\nlp},~\decodec{\vis}) pairs are summarized below (please refer to~\cite{nvbench} for more details).

\stitle{(S1) Synthesizing visualizations.}
It treats an \sql query $Q$ as a tree structure and does tree edits (\eg deleting some tree branches and inserting the type of visualizations), which may result in multiple trees, with each tree corresponding to one possible visualization.

\stitle{(S2) Filtering ``bad'' visualizations.}
In order to ensure that each \vis query is ``good'' (for example, a bar chart with several hundred bars is not readable, and thus is considered to be bad), we need to filter ``bad'' charts. We use a pre-trained machine learning model, namely \deepeye~\cite{deepeye_icde}, to prune synthesized bad \vis queries. \deepeye was trained on 2520/30892 labeled good/bad charts, using features such as the number of distinct values, the number of tuples, the ratio of unique values, max and min values, data type, attribute correlations, and \vis type. Given a \vis query, \deepeye will return either true (\ie a good \vis) or false (\ie a bad \vis).

\stitle{(S3) Synthesizing \nlp queries.}
For the remaining ``good'' visualizations, we need to modify the input \nlp query for \sql (\eg $n_Q$ in Figure~\ref{fig:synthesizer}) to reflect the changes \wrt tree edits, which might result in multiple output \nlp queries, \eg $n_{11}$ (resp. $n_{12}$) is synthesized from $n_Q$ based on the differences between $t_1$ (resp. $t_2$) and $n_Q$. 
For those cases of deleting some parts of the \nlp query for \sql to produce the \nlp query for \vis, \eg $n_{11}$ in Figure~\ref{fig:synthesizer}, we need to interact with the user to produce the \nlp query for \vis.

\stitle{(S4) Manual verification.} 
We asked 23 experts and 312 crowd-workers to verify the quality of synthesized (\encodec{\nlp}, \decodec{\vis}) pairs. Experts/crowd-workers consider 86.9\%/88.7\% of synthesized (\encodec{\nlp}, \decodec{\vis}) pairs are well-matched, \ie scored 4 or 5 in a range $[1, 5]$ with 1 for bad matches and 5 for perfect matches.
As measured by~\cite{nvbench}, our synthesis method reduces the man-hour to 5.7\% of developing an \nlvis benchmark from scratch. In other words, building an \nlvis benchmark by humans 17.5$\times$ man-hours of our method.

%
%
%
%
%
%
%
%


\begin{figure}[t!]
	\centering
	\includegraphics[width=.93\columnwidth]{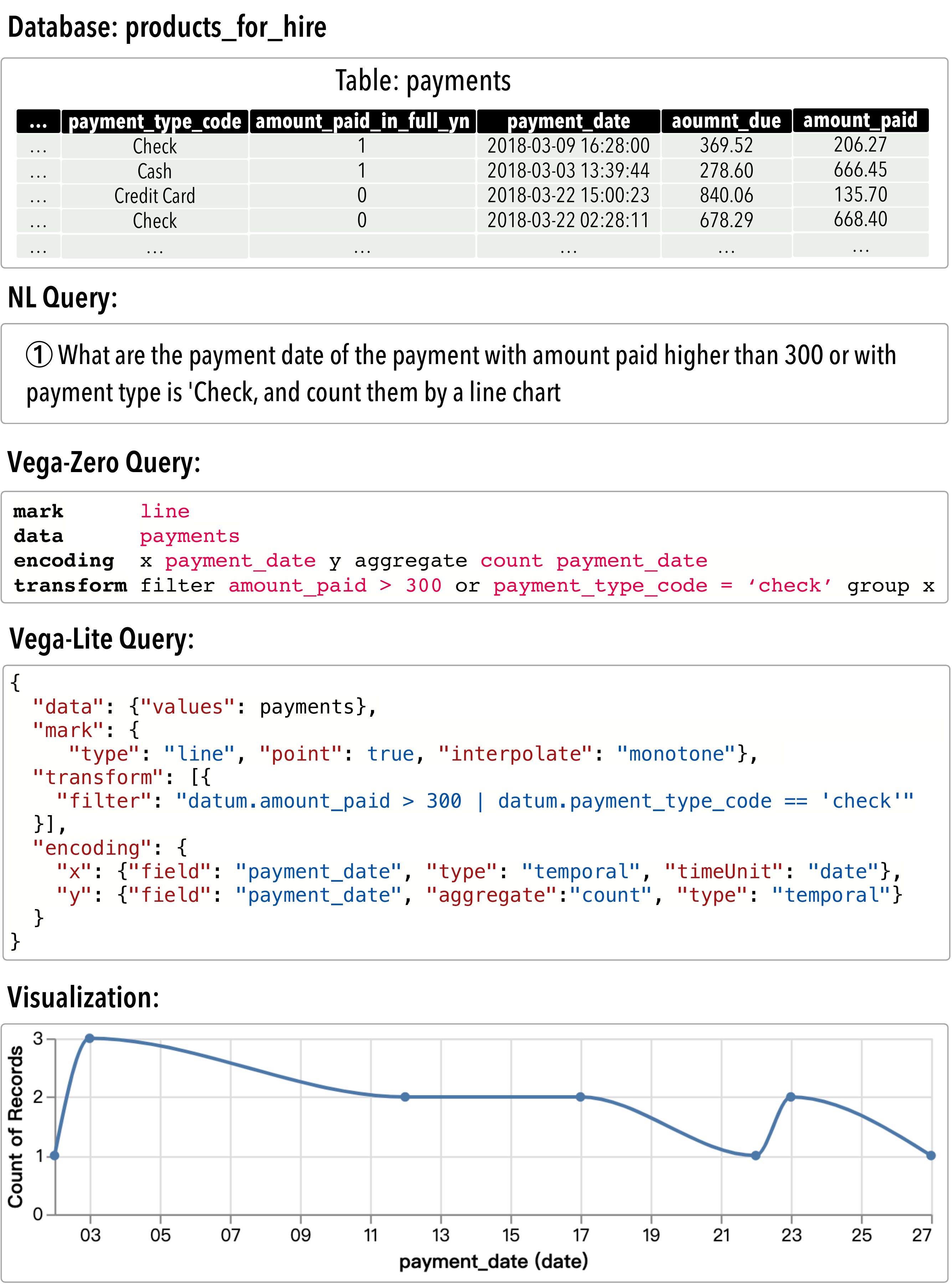}	
	\vspace{-.5em}
	\caption{ An example with filtering operation in \tvbench}
	\vspace{-1em}
	\label{fig:examples}
\end{figure}

\begin{figure}[h!]
	\includegraphics[width=0.98\columnwidth]{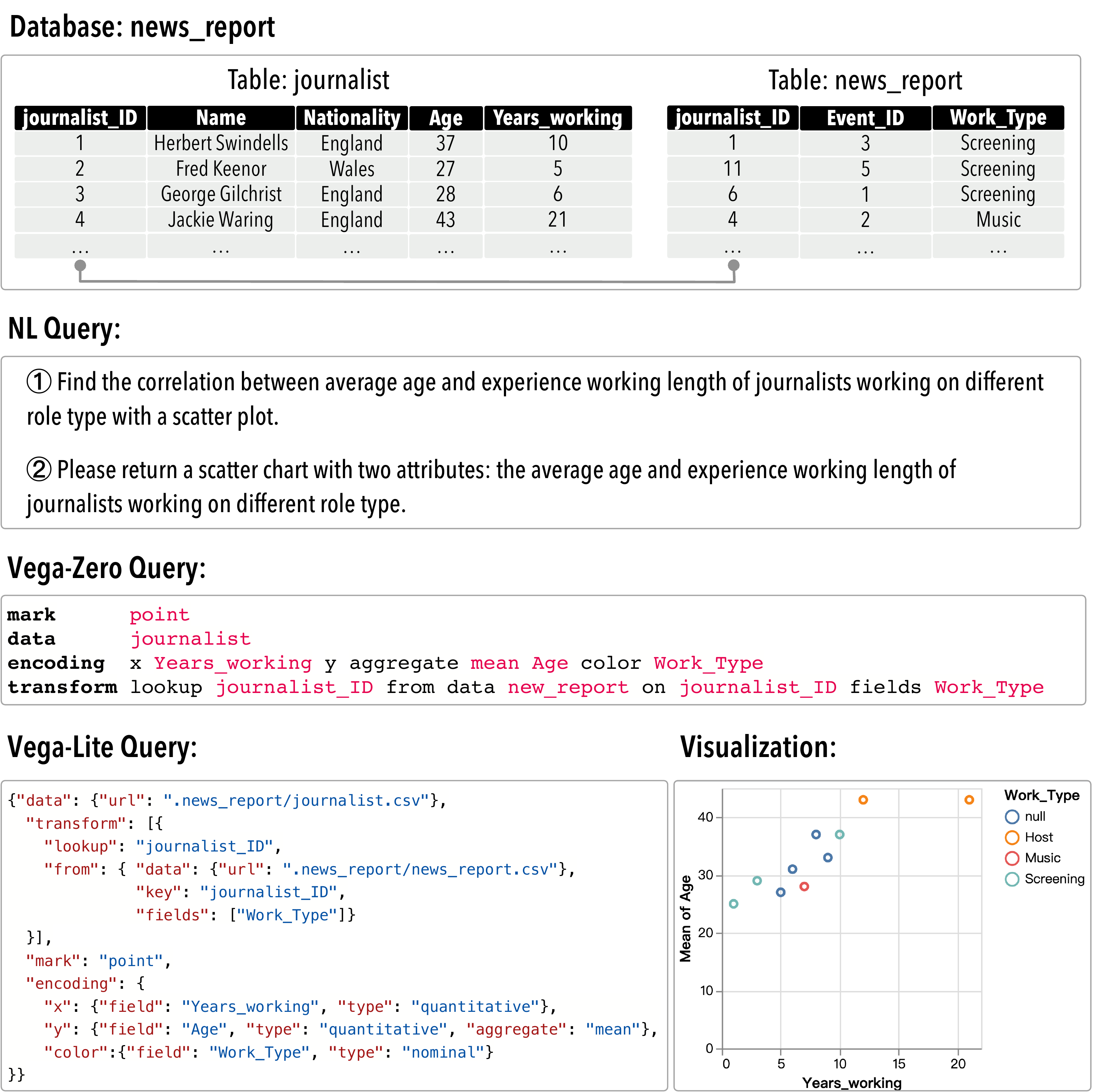}	
	\vspace{-1em}
	\caption{ An example with joining operation in \tvbench}
	\label{fig:joining_examples}
\end{figure}


\section{nvBench: Details}
\label{sec:benchmark}

Figure~\ref{fig:nvbench_stat} overviews the statistics of \tvbench, synthesized from a cross-domain \nlsql benchmark Spider~\cite{DBLP:conf/emnlp/YuZYYWLMLYRZR18}.

\tvbench has 153 databases along with 780 tables in total and covers 105 domains (\eg finance, college). 
Among the columns, 68.78\% of columns are categorical columns, 11.58\% of columns are temporal columns, and 19.64\% of columns are quantitative columns. The maximum number of rows in a table is 183,978, and the minimum number of rows is 1, with an average of 1309.65 rows.

On top of 153 databases, \tvbench contains 7,274 visualizations on seven types of charts. 
For each visualization, \tvbench provides one to several \nlp queries. 
In total, \tvbench consists of 25,750 (\encodec{\nlp},~\decodec{\vis}) pairs. 
For example,
Figure~\ref{fig:examples} shows a (\encodec{\nlp},~\decodec{\vis}) pair in the \tvbench.  The {\vis} query can be presented as the tree format as introduced in \cite{nvbench} or as the Vega-Zero (a Vega-Lite like language proposed in \cite{ncnet}). 
Figure~\ref{fig:joining_examples} showcases a (\encodec{\nlp},~\decodec{\vis}) pair with a more complex operation, \ie joining the data from multiple tables.

\section{Developing NL2VIS Model Using nvBench}
\label{sec:results}

\begin{figure}[t!]
	\hspace{-1.3em}
	\includegraphics[width=1.1\columnwidth]{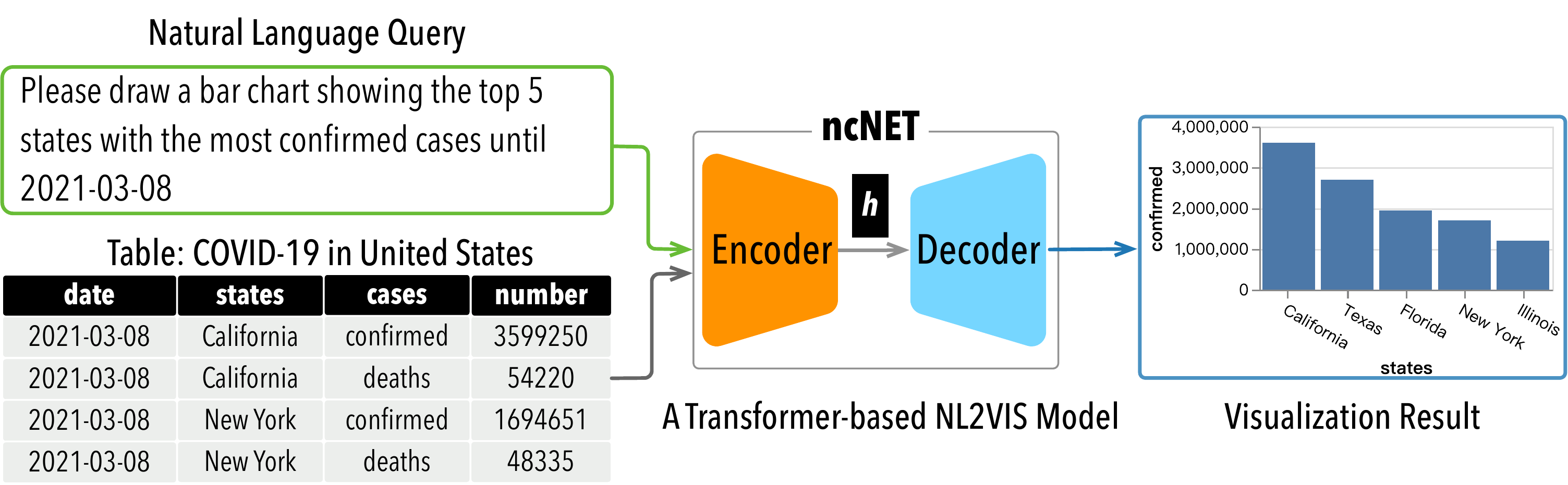}	
	\vspace{-2em}
	\caption{\sys: a Transformer-based \seqseq model for \nlvis}
	\label{fig:seq2vis}
\end{figure}

Given such a large-scale \nlvis benchmark, some exemplary applications of \tvbench including: 
(1) developers can analyze the character of 25,750 \nlp queries to derive some frequent phrases or keywords to help the design of \nlvis interface (\eg \nlp query auto-completion, \nlp phrases suggestion); and
(2) developers can use the \tvbench to train deep learning-based models for the cross-domain \nlvis task. 
Next, we describe more details about how to train a deep learning model using \tvbench.

\stitle{\sys: a Transformer-based \seqseq Model for NL2VIS.}
To learn the translation of  \nlp queries to  \vis queries, one straightforward solution is applying a sequence-to-sequence (\seqseq) model~\cite{DBLP:conf/nips/SutskeverVL14}, similar to translating English to Chinese.
As shown in Figure~\ref{fig:seq2vis}, \sys devises a Transformer-based~\cite{transformer} \seqseq model that consists of two parts, an encoder and a decoder, where each part stacks of self-attention blocks. The task of an encoder is to understand the input sequence, and generate a smaller representation $h$ (\ie a high-dimensional vector) to represent the input. 
The task of a decoder is to generate a sequence of outputs by taking $h$ as input. The \sys needs to be trained with a lot of training data, in the form of (\encodec{Input~sequence},~\decodec{Output~sequence}) pairs.
For example,
a sample \nlp query  is: \textcolor{black!75}{{\tt \textbf{draw a line chart to show the trend of cases number by each case type in Utah}}}, and its corresponding output sequence in a Vega-Lite like language is:

\vspace{0.5em}
\noindent\fbox{%
	\parbox{0.97\columnwidth}{%
		{\tt \textbf{mark} \textcolor[RGB]{0, 113, 188}{\textit{line}} \textbf{encoding} x 
			\textcolor[RGB]{0, 113, 188}{\textit{date}} 
			y aggregate 
			\textcolor[RGB]{0, 113, 188}{\textit{none}}  \textcolor[RGB]{0, 113, 188}{\textit{number}} 
			color \textcolor[RGB]{0, 113, 188}{\textit{cases}} 
			\textbf{transform} filter states = } `{\tt \textcolor[RGB]{0, 113, 188}{\textit{Utah}}}'
	}%
}\vspace{0.5em}

For \nlvis, we train \sys with a lot of (\encodec{\nlp}, \decodec{\vis}) pairs from \tvbench, such that it learns to translate from an \nlp query  to a \vis query.

\begin{figure}[t!]
	\includegraphics[width=\columnwidth]{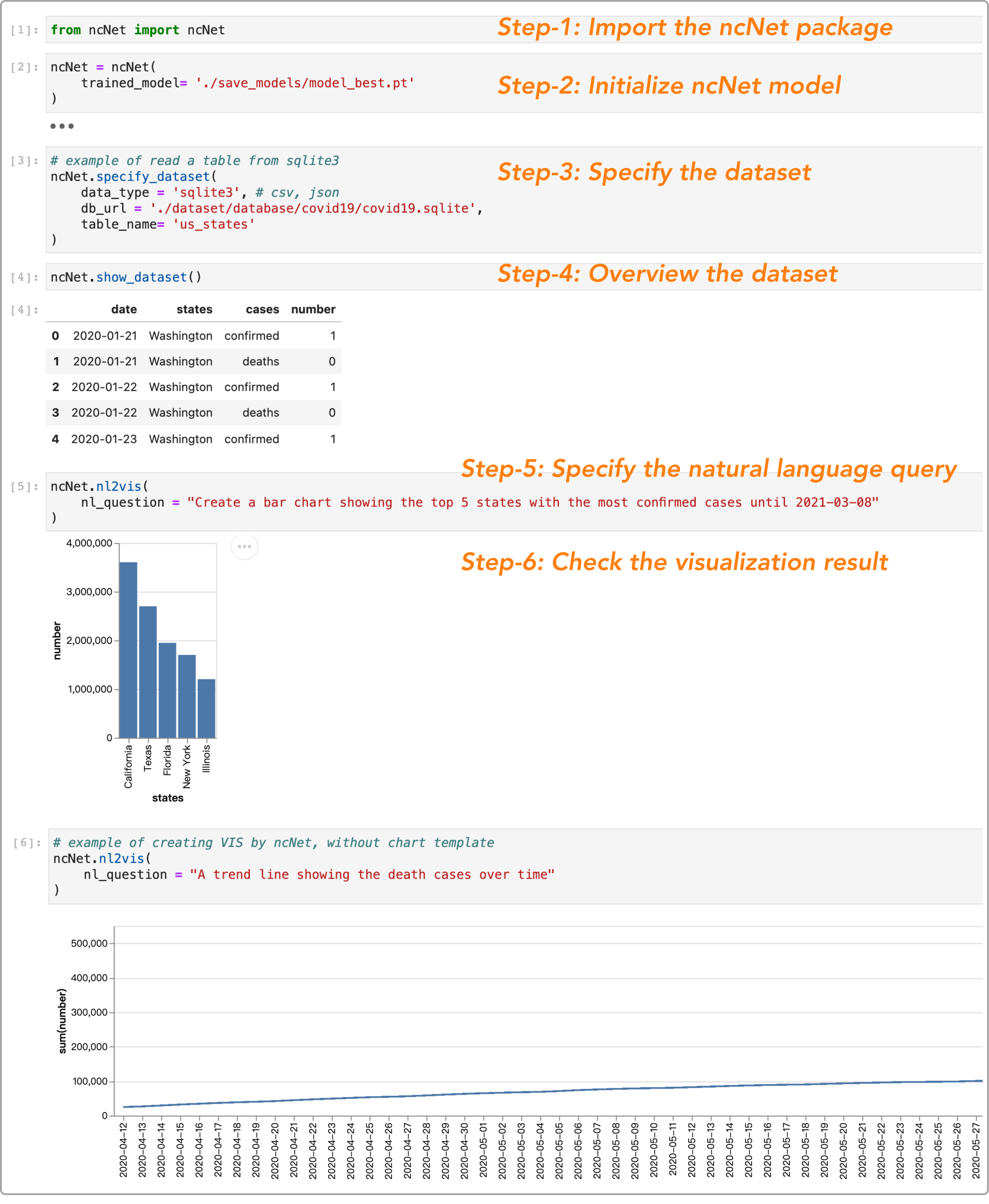}
	\vspace{-2em}
	\caption{Demonstration of using \sys in the Jupyter Lab. }
	\label{fig:usecases}
	\vspace{-.5em}
\end{figure}

\stitle{COVID-19 Use Cases.}
We use a COVID-19 dataset, with the four attributes ({\tt date, states, cases, number}), to demonstrate how the user creates their desired visualization using \nlp query in the Jupyter Lab environment.
We invited data visualization enthusiast Kevin who has experience in building a COVID-19 dashboard.
As shown in Figure~\ref{fig:usecases}, Kevin first  imports the \sys's Python package and then initializes the \sys by passing the model parameter.
Next, he can specify and overview a dataset by calling the function \colorbox{black!20}{{\att{\small specify\_dataset(~)}}}, \colorbox{black!20}{{\att{\small show\_dataset(~)}}}, respectively.
Alternatively, he can explore the dataset using other packages such as {\tt Pandas-profiling}.
In the \nlvis step, 
Kevin specifies an \nlp query via the function \colorbox{black!20}{{\att{\small nl2vis(nl\_question)}}}, and then he can check the visualization given by \sys. If he does not satisfy with the result, he can rephrase the \nlp query and try again.
Before, he spends hours transforming the data and writing Vega-Lite code to visualize; now, Kevin blinks and it’s done.


\section{NL2VIS Benchmark: Where to GO?}
\label{sec:directions}

%
%


There is no doubt that \nlvis benchmarks play a significant role in spawning the boom in the study of \nlvis. 
To make the performance of \nlvis more powerful and robust in real tasks and users, the \nlvis benchmark should cover more diversified tasks, datasets, data types, different characters of \nlp queries, and visualization types.



%


\stitle{Supporting conversational NL2VIS.} 
In real-world visual data analysis scenarios, data analysts usually perform data visualization in a conversational way, \ie conversational visual analysis. 
One conversational \nlp query may consist of a series of standalone but relevant \nlp queries.
Thus, how to extend the \nlvis benchmark to support conversational visual analysis is an interesting and promising direction.

\stitle{Support underspecified \nlp queries.}
	In this work, we assume that an \nlp query $n_V$ can be translated to a valid \vis query $V$, which is based on the assumption that $n_V$ is well specified.
	In practice, $n_V$ could be underspecified, \ie some information to complete $V$ is not provided. 
	From the NLP perspective, this links to the problem of \nlp query auto-completion~\cite{DBLP:conf/naacl/TrnkaYMMP07}.
	From the \vis query perspective, this relates to visualization recommendation~\cite{DBLP:conf/chi/HuBLKH19,DBLP:journals/sigmod/VartakHSMP16}.
	Supporting underspecified \nlp queries is quite straightforward based on our proposal. 
	When translating an \nlp query to a partial \vis tree, it just needs to complete the partial \vis tree to get many valid \vis trees and then rank them using existing works~\cite{DBLP:journals/tvcg/MoritzWNLSHH19,deepeye_icde}.


\stitle{Support more visualization types.}
Currently, \tvbench only consists of seven popular visualization types. 
Future studies can collect (\encodec{\nlp},~\decodec{\vis}) pairs for other popular visualization types, \eg heatmap and box-plot,  to enrich \nlvis benchmarks to cover more diversified tasks. 
Besides, if some cases, \eg mixing two charts in a visualization and visualizations with advanced calculations, can be covered, more practical analysis tasks can be available.

\stitle{Supporting domain-specified NL2VIS.} 
Some domains, \eg chemistry, biology, and healthcare, have their own data structures, data formats, terminology abbreviations, and special phrases of \nlp queries. How to extend  \nlvis benchmarks to support the \nlvis task in these domains is important and interesting.

\stitle{Collecting and characterizing \nlp queries.} 
Beyond the space of \nlp queries and \vis covering in the \nlvis benchmarks, we also need to understand how the real users express their \nlp queries in different visual analysis tasks, domains, and scenarios. 
Srinivasan~et~al.~\cite{nl4vis} have taken the first step towards this goal, but the total amount of \nlp queries they collect and analyze is still relatively small. The study of \nlvis is eager to more and more samples generated by real users and tasks from the mainstream visualization vendors such as Tableau's~Ask~Data~\cite{askdata}.

\stitle{Make it fully automatic}. As shown in Section~\ref{label: nvbench}-\textbf{(S3)}, the only part that is not automatic is to edit the \nlp query for tree deletions. The main challenge is to identify the part of \nlp corresponding to the deletions. This is doable by training a deep learning model that takes tree edits as input and \nlp edits as output if we have enough training data, or use some powerful language models (\eg GPT-3~\cite{gpt3}). 
Thus, if we make the above step a success,  naturally, we can  synthesize as many good (\encodec{\nlp}, \decodec{\vis}) pairs as possible based on the rich collection of \nlsql benchmarks.

\section{Concluding Remarks.}
\label{sec:conclusion}

In this paper, we have introduced \tvbench, the first large-scale \nlvis benchmark that was developed to empower deep learning-based neural machine 
translation  for cross-domain \nlvis task.
We have discussed  how to synthesize \tvbench by piggybacking \nlsql benchmarks. 
We have presented the statistics information about \tvbench and showcased some concrete examples in \tvbench. The quality of \tvbench has been validated by both experts and crowd-workers.
We have also introduced how to train a deep learning-based model for learning the \nlvis translation. 
Our use cases show that \sys, trained using \tvbench, can work well in the \nlvis task. 
We have also outlined some interesting directions about the development of  \nlvis benchmarks, to push the field of \nlvis to its real-world applications.


\vspace{.5em}
\acknowledgments{
	This project is supported by NSF of China (61925205, 61632016), Beijing National Research Center for Information Science and Technology (BNRist), Huawei, TAL Education, and Zhejiang Lab’s International Talent Fund for Young Professionals.
}

\newpage
\balance
\bibliographystyle{abbrv-doi-hyperref}

\bibliography{main}
\end{document}